\documentclass[useAMS,longauth,usenatbib,usegraphicx,traditabstract]{aa}

\usepackage{graphicx}
\usepackage{natbib}
\usepackage{amsmath}
\usepackage{psfig}
\bibpunct{(}{)}{;}{a}{}{,}

\def\aap{A\&A}

\def\apjl{ApJ}
\def\apjs{ApJS}

\def\oiii{[O\,{\sc iii}]}
\def\cii{[C\,{\sc ii}]}
\def\nii{[N\,{\sc ii}]}
\def\oi{[O\,{\sc i}]}

\def\td{$T_{\rm d}$}
\def\gs{\mathrel{\raise0.35ex\hbox{$\scriptstyle >$}\kern-0.6em
\lower0.40ex\hbox{{$\scriptstyle \sim$}}}}
\def\ls{\mathrel{\raise0.35ex\hbox{$\scriptstyle <$}\kern-0.6em
\lower0.40ex\hbox{{$\scriptstyle \sim$}}}}

\begin{document}

\title{{\it Herschel}\thanks{{\it Herschel} is an ESA space
    observatory with science instruments provided by European-led
    Principal Investigator consortia and with important participation
    from NASA.} and SCUBA-2 imaging and spectroscopy of a bright, lensed
    submillimetre galaxy at $z=2.3$}

\titlerunning{FTS spectroscopy of a lensed SMG}

\author{R.\,J.~Ivison\inst{1,2}
\and
A.\,M.~Swinbank\inst{3}
\and
B.~Swinyard\inst{4}
\and
Ian Smail\inst{3}
\and
C.\,P.~Pearson\inst{4,5}
\and
D.~Rigopoulou\inst{4,6}
\and
E.~Polehampton\inst{4,5}
\and
J.-P.~Baluteau\inst{7}
\and
M.\,J.~Barlow\inst{8}
\and
A.\,W.~Blain\inst{9}
\and
J.~Bock\inst{9,10}
\and
D.\,L.~Clements\inst{11}
\and
K.~Coppin\inst{3}
\and
A.~Cooray\inst{12}
\and
A.~Danielson\inst{3}
\and
E.~Dwek\inst{13}
\and
A.\,C.~Edge\inst{3}
\and
A.~Franceschini\inst{14}
\and
T.~Fulton\inst{15}
\and
J.~Glenn\inst{16}
\and
M.~Griffin\inst{17}
\and
K.~Isaak\inst{17}
\and
S.~Leeks\inst{4}
\and
T.~Lim\inst{4}
\and
D.~Naylor\inst{5}
\and
S.\,J.~Oliver\inst{18}
\and
M.\,J.~Page\inst{19}
\and
I.~P{\'e}rez-Fournon\inst{20,21}
\and
M.~Rowan- -Robinson\inst{10}
\and
G.~Savini\inst{22}
\and
D.~Scott\inst{23}
\and
L.~Spencer\inst{17}
\and
I.~Valtchanov\inst{24}
\and
L.~Vigroux\inst{25}
\and
G.\,S.~Wright\inst{1}
}

\authorrunning{Ivison et al.}


\institute{UK Astronomy Technology Centre, Royal Observatory, Blackford Hill,
           Edinburgh EH9 3HJ, UK
\and
           Institute for Astronomy, University of Edinburgh, Royal Observatory,
           Blackford Hill, Edinburgh EH9 3HJ, UK
\and
           Institute for Computational Cosmology, Durham University,
           South Road, Durham DH1 3LE, UK
\and
           Space Science \& Technology Department, Rutherford Appleton Laboratory,
           Chilton, Didcot OX11 0QX, UK
\and
           Institute for Space Imaging Science, University of Lethbridge,
           Lethbridge, Alberta T1K 3M4, Canada
\and
           Astrophysics, Oxford University, Keble Road, Oxford, OX1 3RH, UK
\and
           Observatoire Astronomique de Marseille-Provence, 2 Pl Le Verrier,
           FR 13248, Marseille, Cedex 04, France
\and
           Department of Physics and Astronomy, University College London, 
           Gower Street, London WC1E 6BT, UK
\and
           California Institute of Technology, 1200 E.\ California Blvd,
           Pasadena, CA 91125, USA
\and
           Jet Propulsion Laboratory, Pasadena, California 91109-8099, USA
\and
           Astrophysics Group, Imperial College, Blackett Laboratory,
           Prince Consort Road, London SW7 2AZ, UK
\and
           Center for Cosmology, Department of Physics and Astronomy,
           University of California, Irvine, CA 92697, USA
\and
           Observational Cosmology Laboratory, Code 665, NASA Goddard Space
           Flight Center, Greenbelt, MD 20771, USA
\and
           Dipartimento di Astronomia, Universita' di Padova, vic.\ Osservatorio, 3,
           35122 Padova, Italy
\and
           Blue Sky Spectroscopy, Lethbridge, Alberta, Canada
\and
           Department of Astrophysical and Planetary Sciences, CASA 389-UCB,
           University of Colorado, Boulder, CO 80309, USA
\and
           Cardiff School of Physics and Astronomy, Cardiff University,
           Queens Buildings, The Parade, Cardiff CF24 3AA, UK
\and
           Astronomy Centre, Department of Physics \& Astronomy,
           University of Sussex, Falmer, East Sussex BN1 9QH, UK
\and
           Mullard Space Science Laboratory, University College London,
           Holmbury St Mary, Dorking, Surrey RH5 6NT, UK
\and
           Instituto de Astrof{\'\i}sica de Canarias (IAC), E-38200 La Laguna,
           Tenerife, Spain
\and
           Departamento de Astrof{\'\i}sica, Universidad de La Laguna (ULL),
           E-38205 La Laguna, Tenerife, Spain
\and
           Department of Engineering, University of Cambridge, Cambridge
           CB3 0FA, UK
\and
           Department of Physics \& Astronomy, University of British
           Columbia, 6224 Agricultural Road, Vancouver, BC V6T 1Z1, Canada
\and
           European Space Astronomy Centre, P.O.\ Box 78, 28691 Villanueva
           de la Ca\~nada, Madrid, Spain
\and
           Institut d'Astrophysique de Paris, 98bis, bd Arago --
           75014 Paris, France
}

\date{Received \dots / Accepted \dots}

\abstract{We present a detailed analysis of the far-infrared (-IR)
properties of the bright, lensed, $z=2.3$, submillimetre-selected
galaxy (SMG), SMM\,J2135$-$0102 (hereafter SMM\,J2135), using new
observations with {\it Herschel}, SCUBA-2 and the Very Large Array
(VLA). These data allow us to constrain the galaxy's spectral energy
distribution (SED) and show that it has an intrinsic rest-frame
8--1000-$\mu$m luminosity, $L_{\rm bol}$, of $(2.3\pm0.2)\times
10^{12}$\,L$_{\odot}$ and a likely star-formation rate (SFR) of $\sim
400$\,M$_{\odot}$\,yr$^{-1}$. The galaxy sits on the far-IR/radio
correlation for far-IR-selected galaxies. At $\gs 70$\,$\mu$m, the SED
can be described adequately by dust components with dust temperatures,
\td\ $\sim$ 30 and 60\,{\sc k}.  Using SPIRE's Fourier Transform
Spectrometer (FTS) we report a detection of the \cii\,158$\mu$m
cooling line. If the \cii, CO and far-IR continuum arise in
photo-dissociation regions (PDRs), we derive a characteristic gas
density, $n\sim 10^{3}$\,cm$^{-3}$, and a far-ultraviolet (-UV)
radiation field, $G_0$, $10^3\times$ stronger than the Milky Way.
$L_{\rm [CII]}/L_{\rm bol}$ is significantly higher than in local
ultra-luminous IR galaxies (ULIRGs) but similar to the values found in
local star-forming galaxies and starburst nuclei.  This is consistent
with SMM\,J2135 being powered by starburst clumps distributed across
$\sim$2\,kpc, evidence that SMGs are not simply scaled-up ULIRGs. Our
results show that SPIRE's FTS has the ability to measure the redshifts
of distant, obscured galaxies via the blind detection of atomic
cooling lines, but it will not be competitive with ground-based
CO-line searches. It will, however, allow detailed study of the
integrated properties of high-redshift galaxies, as well as the
chemistry of their interstellar medium (ISM), once more suitably
bright candidates have been found.}


\maketitle

\section{Introduction}

Submillimetre (submm) surveys have uncovered a population of
intrinsically luminous, but highly obscured, galaxies at high
redshift.  However, even with instrinsic luminosities of
$\sim10^{13}$\,L$_{\odot}$ \citep[e.g.][]{ivison98}, the brightest
SMGs are still challenging targets for observational studies. In
the submm and far-IR, where the bulk of their luminosity escapes, the
brightest SMGs have observed flux densities of only $\sim10$\,mJy at
850\,$\mu$m, peaking at $\sim50$\,mJy at the wavelengths probed by
{\it Herschel}. To alleviate this photon starvation, submm surveys
often exploit gravitational lensing via massive, foreground galaxy
clusters, thereby enhancing the apparent brightness of SMGs at all
wavelengths \citep[e.g.][]{smail97, chapman02, cowie02}.

Recently, \citet{swinbank10a} exploited the cluster lensing technique
using the Large Apex Bolometer Camera \citep[LABOCA --][]{siringo09}
on the 12-m Atacama Pathfinder Experiment (APEX) telescope to map the
cluster, MACS\,J2135$-$01 ($z=0.325$), and thereby discovered
SMM\,J2135, an SMG with $S_{\rm 870\mu m} =106$\,mJy. Its brightness
is due to very high amplification (by $32.5\pm4.5$) by the foreground
cluster (similarly bright sources may have recently been unearthed by
the South Pole Telescope -- \citealt{vieira10}).  The lens model for
SMM\,J2135 is well constrained and its redshift ($z=2.3259\pm 0.0001$,
derived from the detection of CO $J$=1--0 in a blind search) and
intrinsic flux ($3.3\pm0.5$\,mJy) are typical of SMGs found close to
the confusion limit in submm surveys. SMM\,J2135 thus presents an
opportunity to study a member of this important population at high
signal-to-noise and with the spatial and spectral resolution necessary
to determine the detailed far-IR spectral properties of SMGs. Due to
the high magnification, it is feasible to apply some of the
observational tools used on local star-forming galaxies to understand
the processes of star formation at high redshift.  Indeed, we can
employ diagnostics capable of determining the flux of ionising
radiation and the SFR, thus determining the state of the overwhelming
majority of the atomic and molecular gas in this galaxy
\citep{wolfire90, hollenbach99, kaufman99}.

In this paper we present spectroscopic and photometric far-IR/submm
measurements of SMM\,J2135 made using {\it Herschel}
\citep{pilbratt10}.  We also include new observations with the James
Clerk Maxwell Telescope (JCMT) and VLA. We use these observations to
constrain the SED of SMM\,J2135 and measure or set firm limits for the
line fluxes from the main atomic cooling lines.

\section{Observations}

To complement the existing submm photometry of SMM\,J2135,
observations at 250, 350 and 500\,$\mu$m were obtained with SPIRE
\citep{griffin10}. The field was observed first using the `small-map
mode', where orthogonal scans produce a useful cross-linked area of
$\sim$16\,arcmin$^2$. We used four repetitions, giving an on-source
integration time of $\sim$200\,s. Processing relied on the SPIRE Scan
Map Pipeline \citep{griffin08}, which deglitches, flux calibrates and
performs various corrections. After removal of a linear baseline,
images were made using the standard naive mapper within the {\it
Herschel} Interactive Pipeline Environment (HIPE~v2.0).  From the
final maps, we identify a $\sim$100-$\sigma$ source at the position
of SMM\,J2135 in all bands; its flux densities are listed in
Table~\ref{tab:Phot}.

SMM\,J2135 was also observed for 7\,ks using the central pixels of
SPIRE's FTS (covering $\lambda_{\rm obs} = 197$--$670\mu$m) on 2009
December 9, to search for \cii\,158$\mu$m, redshifted to
524\,$\mu$m. Even with the benefit of extreme amplification,
SMM\,J2135 represents an extremely faint target in the context of the
SPIRE spectrometer: the standard pipeline reduction shows significant
problems with the overall flux level in both the high- and
low-frequency channels (SSW, SLW). Rather than rely on the pipeline,
we used the variation in bolometer temperature to transform the source
and dark interferograms into spectra which were then subtracted and
divided by a calibration spectrum of Uranus (rather than the much
fainter asteroid, Vesta -- see \citealt{swinyard10}).  Variations in
instrument temperature between the observations of the dark sky and
the source can cause large relative variations in the SLW
spectrum. Here, we determined the overall net flux of the source, with
no subtraction or addition of flux from the variation in instrument
temperature. We then inspected the SLW data and compared to the
spectrum expected from the subtraction of two blackbodies at the
temperatures recorded in the housekeeping data.  The difference in
model instrument temperatures in the dark sky and the source
observation are therefore varied (by less than 1\%) until a match
between the overall flux level from the photometer and SSW is
achieved.

New observations were also carried out with the Submillimetre
Common-User Bolometer Array-2 \citep[SCUBA-2 --][]{holland06}, a
large-format bolometer camera for the JCMT, designed to produce
simultaneous continuum images at 450 and 850\,$\mu$m. These data were
obtained during 2009 November 29, during early commissioning, with one
32$\times$40 transition-edge sensor (TES) array at each of 450 and
850\,$\mu$m, giving a field of view of $\sim 3'\times3.5'$ (the final
commissioned instrument will have four such arrays at each
wavelength). The total integration time was 3.6\,ks. Pointing checks
and flux calibration was achieved via observations of Neptune and
Uranus, immediately before and after the science exposures. Data
reduction was carried out using the Submm User Reduction Facility
({\sc smurf}), which flatfields and stacks the images, and removes
atmospheric emission. Measured flux densities are listed in
Table~\ref{tab:Phot}.

To determine the radio properties of the galaxy, observations with the
VLA were obtained during late 2009.  SMM\,J2135 was observed in the C
and X bands for 10 and 5\,ks, respectively. The C-band observations
were taken in spectral-line mode, to search for redshifted 22-GHz
water maser emission, though only continuum was detected; continuum
emission was also detected convincingly in the X band
(Table~\ref{tab:Phot}).

%
%
\begin{figure}
\centerline{
\psfig{figure=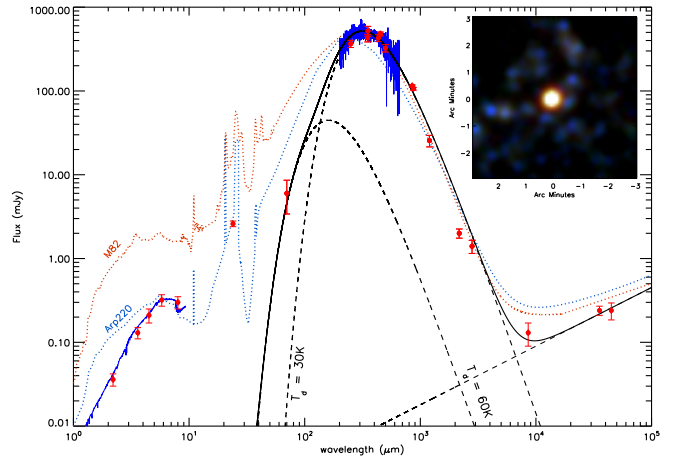,width=3.4in,angle=90}
}
\caption{The rest-frame near-IR--radio SED of SMM\,J2135, with
new {\it Herschel}, SCUBA-2 and VLA observations complementing
existing photometry \citep{swinbank10a}. The FTS spectrum is shown in
blue. In the rest-frame optical to mid-IR regime, SMM\,J2135 is less
luminous than Arp\,220 and considerably fainter than M\,82, possibly
reflecting strong dust obscuration. We model the SED using a
two-component dust model (solid, black line) comprising two modified
blackbodies ($\beta=+2.0$) with \td\ = 30 and 60\,{\sc k}. The solid
blue line denotes a stellar fit to the rest-frame UV--near-IR
photometry. Inset is a colour image, centred on SMM\,J2135, generated
from the SPIRE 250-, 350- and 500-$\mu$m observations (N, up; E,
left).}
\label{fig:SED}
\end{figure}

%
%
\begin{figure}
\centerline{
\psfig{figure=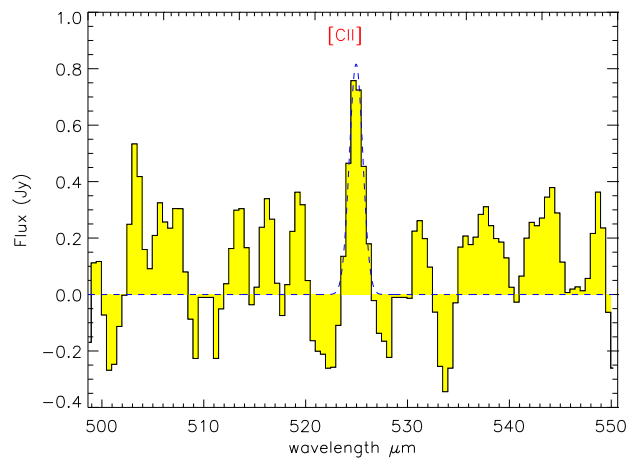,width=3.2in,angle=90}
}
\caption{Region around the redshifted \cii\,158$\mu$m, the strongest
atomic fine-structure line detected by our FTS spectrum of
SMM\,J2135. Dashed line: best Gaussian fit, with $v_{\rm lsr}=-180\pm
150$\,km\,s$^{-1}$, which corresponds to strong components in the
HCN, C\,{\sc i} and CO lines (Danielson et al., in preparation).
Using the line flux and following equation~1 of \citet{hd10}, we
estimate a gas mass, $M_{\rm [CII]}\sim4\times 10^9$\,M$_{\odot}$,
which is $\sim$25\% of the total molecular gas mass, similar to the
ratio found in local starburst galaxies \citep{stacey91}.}
\label{fig:FTS}
\end{figure}

%
%
\begin{figure}
\centerline{
\psfig{figure=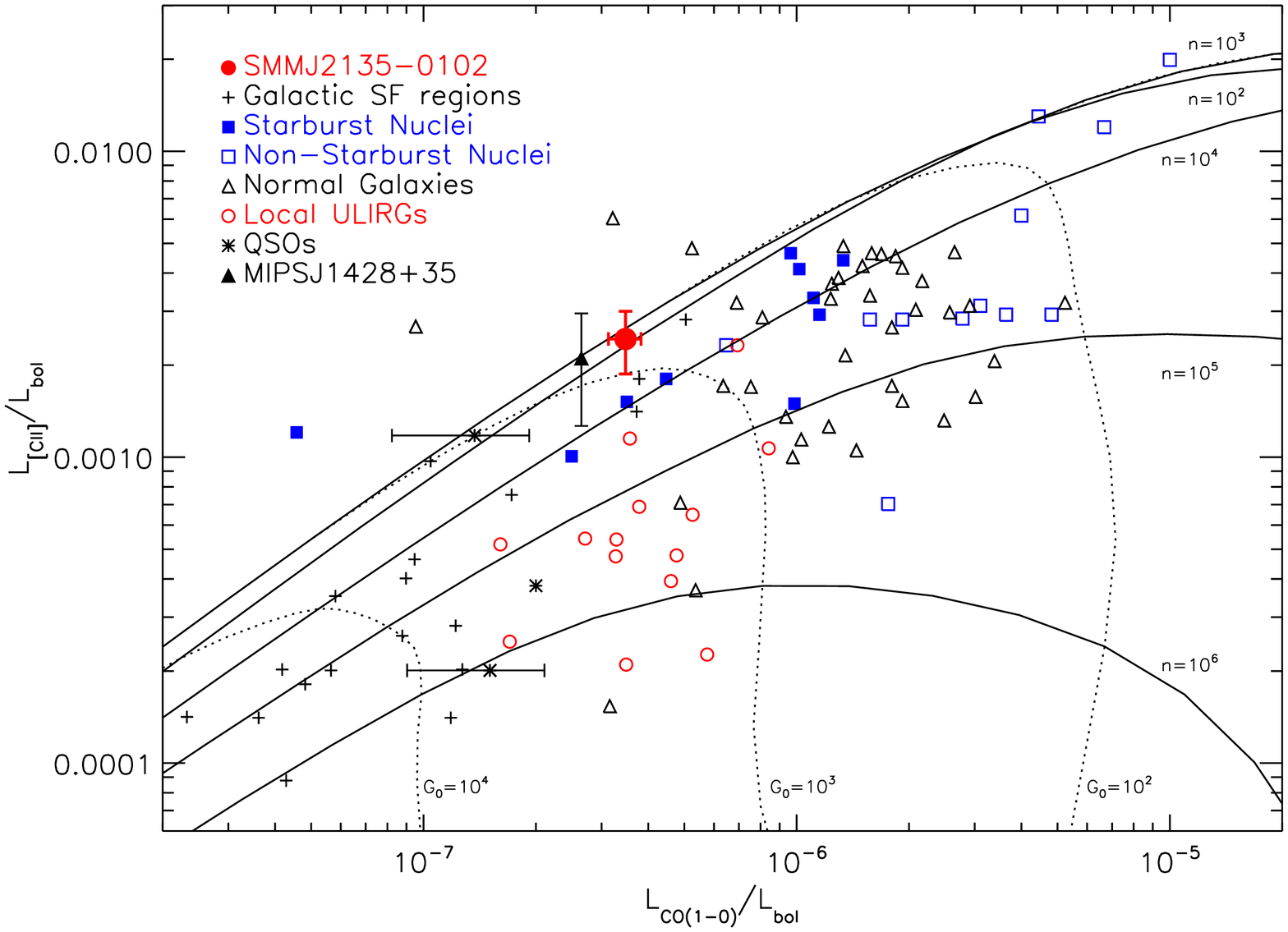,width=3.7in,angle=90}
}
\caption{$L_{\rm [CII]}/L_{\rm bol}$ versus $L_{\rm CO(1-0)}/L_{\rm
bol}$ for SMM\,J2135 compared to star-forming regions, star-forming
galaxies and ULIRGs in the local Universe.  This figure is adapted from
\citet{hd10} and shows the ratios for powerful, high-redshift QSOs as
well as the SMG, MIPS\,J1428+35. Tracks for PDR models of
gas density, $n$, and far-UV field strength, $G_0$, are taken from
\citet{kaufman99}.  We see that the gas in SMM\,J2135 experiences a
far-UV field similar to that seen in local ULIRGs, but is at much
lower densities than the typical material in such systems.}
\label{fig:LCII}
\end{figure}

%
%
\begin{table}
\begin{minipage}[t]{\columnwidth}
\caption{Photometry}
\label{tab:Phot}
\centering
\renewcommand{\footnoterule}{}
\begin{tabular}{lccc}
\hline \hline
\noalign{\smallskip}
Wavelength & Flux\footnote{Errors include uncertainty in absolute flux calibration.} (mJy)     & Observatory/Instrument       \\
 \hline
250\,$\mu$m         & 366\,$\pm$\,55  & {\it Herschel}/SPIRE \\
350\,$\mu$m         & 429\,$\pm$\,64  & {\it Herschel}/SPIRE \\
352\,$\mu$m         & 520\,$\pm$\,70  & APEX/SABOCA\footnote{See \citet{swinbank10a}, also for $\lambda_{\rm obs}<250\,\mu$m.}\\
434\,$\mu$m         & 430\,$\pm$\,40  & SMA$^b$ \\
450\,$\mu$m         & 480\,$\pm$\,54  & SCUBA-2 \\
500\,$\mu$m         & 325\,$\pm$\,49  & {\it Herschel}/SPIRE \\
850\,$\mu$m         & 115\,$\pm$\,13   & SCUBA-2 \\
870\,$\mu$m         & 106\,$\pm$\,12   & APEX/LABOCA$^b$ \\
1.2\,mm	            & 26\,$\pm$\,4    & SMA$^b$ \\
2.17\,mm	    & 2.0\,$\pm$\,0.25    & PdBI$^b$ \\
2.80\,mm	    & 1.4\,$\pm$\,0.25    & PdBI$^b$ \\
8.57\,mm	    & 0.13\,$\pm$\,0.05   & GBT/Zpectrometer$^b$ \\
3.55\,cm	    & 0.240\,$\pm$\,0.030 & VLA/X \\
4.49\,cm	    & 0.240\,$\pm$\,0.055 & VLA/C \\
\hline
\end{tabular}
\end{minipage}
\end{table}

%
%
\begin{table}
\begin{minipage}[t]{\columnwidth}
\caption{Spectral-line and bolometric luminosities.}
\label{tab:linelum}
\centering
\renewcommand{\footnoterule}{}
\begin{tabular}{lccc}
\hline\hline
\noalign{\smallskip}
Line        & $\lambda_{\rm rest}$  & Flux\footnote{Line width constrained to instrumental resolution.}&Luminosity\\
            & ($\mu$m)              & ($\times10^{-17}$\,W\,m$^{-2}$)  & (L$_{\odot}$) \\
\hline
\oi  & 63.18  & $3\sigma<4.5$ & $3\sigma<14.9\times 10^9$\\
\oiii& 88.36  & $3\sigma<2.4$ & $3\sigma<8.0\times 10^9$\\
\nii & 122.10 & $3\sigma<1.4$ & $3\sigma<4.7\times 10^9$ \\
\oi  & 145.53 & $3\sigma<2.5$ & $3\sigma<8.5\times 10^9$ \\
\cii & 157.74 & $1.7\pm 0.4$  & $(5.5\pm 1.3)\times 10^9$ \\
CO(1--0)& 2602.6& $2.14\pm 0.12$\footnote{Jy\,km\,s$^{-1}$.}&$(8.0\pm 0.4)\times 10^5$ \\
$L_{\rm bol}$ & ...    & ...          & $(2.3\pm 0.2)\times 10^{12}$\\
\hline
\end{tabular}
\end{minipage}
\end{table}

\section{Analysis and discussion}

\subsection{Far-infrared SED}

The new observations clearly identify a turnover in the SED of
SMM\,J2135 at $\sim$350\,$\mu$m (Fig.~\ref{fig:SED}). We use the
far-IR photometry (Table~\ref{tab:Phot} and \citealt{swinbank10a}) to
calculate its rest-frame 8--1000-$\mu$m luminosity directly, which is
due largely to dust-reprocessed UV light and provides a measure of its
instantaneous SFR. Correcting for lensing amplification, we find
$L_{\rm bol}=(2.3\pm 0.2)\times 10^{12}$\,L$_{\odot}$, indicating a
SFR of $\sim$400\,M$_{\odot}$\,yr$^{-1}$ \citep{kennicutt98a}. $L_{\rm
bol}$ is thus comparable to that of Arp\,220 and rather higher than
that quoted by \citet{swinbank10a} who integrated the best modified
blackbody fit to the 350-, 434- and 870-$\mu$m emission, missing much
of the energy at rest-frame $\sim$8--100\,$\mu$m.

If we parameterise the far-IR SED of SMM\,J2135 using a modified
blackbody spectrum, a single component model with \td\ = 34\,{\sc k}
underestimates $S_{\rm 70\mu m}$ by $\sim$100$\times$. A two-component
model with \td\ = 30 and 60\,{\sc k} provides a significantly improved
fit (Fig.~\ref{fig:SED}). The mass of dust associated with the warm
and cool components are $M^{\rm warm}_{\rm d}=10^{6}$ and $M^{\rm
  cool}_{\rm d}=4\times 10^{8}$\,M$_{\odot}$ \citep[adopting the
  parameters used by][]{dunne00}. Given the cold molecular gas mass
derived from the CO(1--0) emission ($M_{\rm gas}=(16\pm
1)\times10^{9}$\,M$_{\odot}$ -- \citealt{swinbank10a}), this suggests
a gas-to-dust ratio of $M_{\rm gas}/M_{\rm d}\sim40$, rather lower
than that of the Milky Way, 120, and Lyman-break galaxies
\citep[$\sim$100; e.g.][]{coppin07} but consistent with typical SMGs
\citep[$\sim$60; e.g.][]{coppin08} given that the uncertainties are
considerable.

\subsection{Radio properties}

If the radio spectrum of SMM\,J2135 follows a $S_{\nu}\propto
\nu^{-0.7}$ power law, which is consistent with the data but by no
means certain (Fig.~\ref{fig:SED}; Table~\ref{tab:Phot}), then its
radio luminosity is $L_{\rm 1.4GHz}=9\times 10^{23}$\,W\,Hz$^{-1}$ so
that $q_{\rm IR}=2.42\pm 0.06$, entirely consistent with the
far-IR/radio correlation for 250-$\mu$m-selected galaxies
\citep[$<\!\!q_{\rm IR}\!\!> = 2.40$ --][]{ivison10a}.

\subsection{Spectral properties}

The full FTS spectrum (Fig.~\ref{fig:SED}) covers the major
fine-structure cooling lines and we detect one strong emission line,
\cii$\lambda$158\,$\mu$m, at the 4.3-$\sigma$ level
(Fig.~\ref{fig:FTS}). Table~\ref{tab:linelum} presents the best-fit
flux with the width constrained to the instrumental resolution. The
flux is not sensitive to the fit parameters, for example returning
values well within 1\,$\sigma$ for a line fixed at $v_{\rm
lsr}=0$\,km\,s$^{-1}$. The FTS spectrum covers several other lines and
although we see hints of emission associated with
\oi$\lambda$145\,$\mu$m and \nii$\lambda$122\,$\mu$m, we have chosen
to report conservative upper limits (best-bet flux plus 3$\sigma$) on
these and other lines in Table~\ref{tab:linelum}.

\cii\ is one of the brightest emission lines in star-forming galaxies,
typically accounting for 0.1--1\% of $L_{\rm bol}$. It arises from the
warm and dense PDRs that form on the UV-illuminated surfaces of
molecular clouds, though the \cii\ flux from diffuse H\,{\sc ii}
regions or from diffuse PDRs can be considerable
\citep[e.g.][]{madden93, lord96}. In local star-forming galaxies,
$L_{\rm [C\,II]}/L_{\rm bol}$ and $L_{\rm [CII]}/L_{\rm CO(1-0)}$
provide a sensitive test of the physical conditions within the ISM.
For SMM\,J2135 we find $L_{\rm [CII]}/L_{\rm bol} =
(2.4\pm0.6)\times10^{-3}$ and $L_{\rm CO(1-0)}/L_{\rm bol}= (3.5\pm
0.5)\pm\times 10^{-7}$ and compare these to measurements of local
galaxy populations in Fig.~\ref{fig:LCII}.  We see that $L_{\rm
[CII]}/L_{\rm CO(1-0)}$ in SMM\,J2135 is similar to local ULIRGs, but
that $L_{\rm CO(1-0)}/L_{\rm bol}$ is consistent with the ratios found
in more typical star-forming galaxies and nuclei.

The \cii\ transition is a primary PDR coolant and is a sensitive probe
of both the physical conditions of the photo-dissociated gas and the
intensity of the ambient stellar radiation field \citep{hollenbach99}.
Hence using the PDR models of \citet{kaufman99} we can determine an
acceptable range of temperature, $T$, and gas density, $n$, in
SMM\,J2135, from our measurements of \cii, CO(1--0) and $L_{\rm
bol}$. In these models, $L_{\rm [CII]}/L_{\rm CO(1-0)}$ is most
sensitive to $n$ whilst $L_{\rm [CII]}/L_{\rm bol}$ is sensitive to
the incident far-UV field strength, $G_0$, and hence
$T$. Fig.~\ref{fig:LCII} shows $L_{\rm [CII]}/L_{\rm bol}$ versus
$L_{\rm CO1-0}/L_{\rm bol}$ and suggests a best-fit density,
$n\sim10^{3}$\,cm$^{-3}$, with $T\sim400$\,{\sc k} and $G_0\sim10^{3}$
\citep{kaufman99}. $G_0$ is measured in multiples of the local
interstellar value, so the far-UV radiation field illuminating the
PDRs is $\sim 10^3\times$ more intense than that in the Milky Way, but
comparable to that found in local ULIRGs and the $z=1.3$ SMG,
MIPS\,J1428 \citep{hd10}, while the densities in SMM\,J2135 ($n\sim
10^3$) are most similar to those found in normal star-forming
galaxies, 10--100$\times$ lower than those seen in local ULIRGs.

Taken together, this suggests that the molecular emission does not
reside in a single, compact region, illuminated by an intense UV
radiation field, but that the material is more extended, with the high
$L_{\rm [CII]}/L_{\rm bol}$ ratio then reflecting the lower density of
this extended medium.  Indeed, \citet{swinbank10a} show that although
the rest-frame 260-$\mu$m emission is dominated by four star-forming
regions, each $\sim$100\,pc across, the emission extends over
$\sim$2\,kpc. The size of the star-forming region in SMM\,J2135 is
also comparable to the sizes of the dense gas reservoirs inferred from
high-$J$ CO mapping, $\sim$3\,kpc \citep{tacconi08}.  Thus SMM\,J2135
appears to be powered by an intense starburst whose influence is felt
over a larger region than those seen in local ULIRGs, as has been
suggested for SMGs using radio, submm and CO sizes \citep{biggs08,
  younger08, byi10, ivison10b}, near- and mid-IR colours and spectra
\citep{hainline09, md09} and other far-IR spectroscopy \citep{hd10}.

\section{Discussion \& Conclusions}

We have delineated the far-IR SED of a highly magnified (but
intrinsically typical) SMG, SMM\,J2135, at $z=2.3$. Its rest-frame
8--1000-$\mu$m and 1.4-GHz luminosities are $2.3\times
10^{12}$\,L$_{\odot}$ and $9\times 10^{23}$\,W\,Hz$^{-1}$, with SFR
$\sim$ 400\,M$_{\odot}$\,yr$^{-1}$, and it sits on the far-IR/radio
correlation for starburst galaxies.

{\it Herschel} FTS spectroscopy detects the redshifted [C\,{\sc
ii}]\,158\,$\mu$m emission line, allowing us to investigate the
properties of its ISM. The line luminosity suggests that the mass of
\cii\ is $\sim$25\% of the molecular gas, similar to the ratio found
in local starbursts.

We use CO(1--0), \cii\ and $L_{\rm bol}$ to investigate the ISM's
physical conditions.  From a comparison with PDR models, we derive a
far-UV radiation field, $G_0$, which is $\sim10^3\times$ higher than
that in the Milky Way, but comparable to those found in ULIRGs.  In
contrast, we find a characteristic density, $n\sim 10^{3}$\,cm$^{-3}$,
which is lower than seen in ULIRGs, but comparable to values seen in
local star-forming galaxies and nuclei, as well as a small number of
high-redshift systems where similar measurements have been made.
Together these results suggest that SMM\,J2135 has a SFR intensity
similar to that seen in local ULIRGs, but distributed over a larger
volume. This is consistent with the $\sim$2-kpc distribution of star
formation across this galaxy \citep[][]{swinbank10a} and previous
suggestions of extended star formation in SMGs
\citep[e.g.][]{biggs08}.

Our results show that SPIRE's FTS has the ability to measure the
redshifts of suitably bright and distant, obscured galaxies via
detection of atomic cooling lines such as \cii. However, we estimate
that $\gs$10-hr integrations will be required and this is not
competitive with blind, ground-based CO-line searches
\citep[e.g.][]{weiss09}, as evidenced by the ease with which the
redshift of SMM\,J2135 was determined using Zpectrometer on the Green
Bank Telescope \citep{swinbank10a}.  Nevertheless, our results show
that facilities such as {\it Herschel} and SCUBA-2 will allow detailed
study of the integrated properties of high-redshift galaxies (through
SED modelling), as well as the chemistry of their ISM.

\begin{acknowledgements} We thank Steve Hailey-Dunsheath for useful
discussion.  We thank Fred Lo for granting DDT observations, and Wayne
Holland for observing SMM\,J2135 during SCUBA-2 commissioning. SPIRE
has been developed by a consortium of institutes led by Cardiff
Univ.\ (UK) and including Univ.\ Lethbridge (Canada); NAOC (China);
CEA, LAM (France); IFSI, Univ. Padua (Italy); IAC (Spain); Stockholm
Observatory (Sweden); Imperial College London, RAL, UCL-MSSL, UKATC,
Univ.\ Sussex (UK); Caltech, JPL, NHSC, Univ.\ Colorado (USA). This
development has been supported by national funding agencies: CSA
(Canada); NAOC (China); CEA, CNES, CNRS (France); ASI (Italy); MCINN
(Spain); SNSB (Sweden); STFC (UK); and NASA (USA). SCUBA-2 is
funded by STFC, the JCMT Development Fund and the Canadian Foundation
for Innovation.
\end{acknowledgements}

\bibliographystyle{aa}
\bibliography{14548}

\begin{thebibliography}{32}
\expandafter\ifx\csname natexlab\endcsname\relax\def\natexlab#1{#1}\fi

\bibitem[{{Biggs} \& {Ivison}(2008)}]{biggs08}
{Biggs}, A.~D. \& {Ivison}, R.~J. 2008, \mnras, 385, 893

\bibitem[{{Biggs} {et~al.}(2010){Biggs}, {Younger}, \& {Ivison}}]{byi10}
{Biggs}, A.~D., {Younger}, J.~D., \& {Ivison}, R.~J. 2010, arXiv:1004.0009

\bibitem[{{Chapman} {et~al.}(2002){Chapman}, {Scott}, {Borys}, \&
  {Fahlman}}]{chapman02}
{Chapman}, S.~C., {Scott}, D., {Borys}, C., \& {Fahlman}, G.~G. 2002, \mnras,
  330, 92

\bibitem[{{Coppin} {et~al.}(2008){Coppin}, {Swinbank}, {Neri}, {Cox},
  {Alexander}, {Smail}, {Page}, {Stevens}, {Knudsen}, {Ivison}, {Beelen},
  {Bertoldi}, \& {Omont}}]{coppin08}
{Coppin}, K., {Swinbank}, A.~M., {Neri}, R., {et~al.} 2008, \mnras, 389, 45

\bibitem[{{Coppin} {et~al.}(2007){Coppin}, {Swinbank}, {Neri}, {Cox}, {Smail},
  {Ellis}, {Geach}, {Siana}, {Teplitz}, {Dye}, {Kneib}, {Edge}, \&
  {Richard}}]{coppin07}
{Coppin}, K., {Swinbank}, A.~M., {Neri}, R., {et~al.} 2007, \apj, 665, 936

\bibitem[{{Cowie} {et~al.}(2002){Cowie}, {Barger}, \& {Kneib}}]{cowie02}
{Cowie}, L.~L., {Barger}, A.~J., \& {Kneib}, J. 2002, \aj, 123, 2197

\bibitem[{{Dunne} {et~al.}(2000){Dunne}, {Eales}, {Edmunds}, {Ivison},
  {Alexander}, \& {Clements}}]{dunne00}
{Dunne}, L., {Eales}, S., {Edmunds}, M., {et~al.} 2000, \mnras, 315, 115

\bibitem[{{Griffin} {et~al.}(2010){Griffin}, {Abergel}, {Abreu}, {Ade},
  {Agnese}, {Andr\'e}, {Zheng}, {Knudsen}, {Coppin}, {Kovacs}, {Bell}, {de
  Breuck}, {Dannerbauer}, {Dickinson}, {Gawiser}, {Lutz}, {Rix}, {Schinnerer},
  {Alexander}, {Bertoldi}, {Brandt}, {Chapman}, {Ivison}, {Koekemoer},
  {Kreysa}, {Kurczynski}, {Menten}, {Siringo}, {Swinbank}, \& {van der
  Werf}}]{griffin10}
{Griffin}, M., {Abergel}, A., {Abreu}, A., {et~al.} 2010, \aap, this volume

\bibitem[{{Griffin} {et~al.}(2008){Griffin}, {Dowell}, {Lim}, {Bendo}, {Bock},
  {Cara}, {Castro-Rodriguez}, {Chanial}, {Clements}, {Gastaud}, {Guest},
  {Glenn}, {Hristov}, {King}, {Laurent}, {Lu}, {Mainetti}, {Morris}, {Nguyen},
  {Panuzzo}, {Pearson}, {Pinsard}, {Pohlen}, {Polehampton}, {Rizzo}, {Schulz},
  {Schwartz}, {Sibthorpe}, {Swinyard}, {Xu}, \& {Zhang}}]{griffin08}
{Griffin}, M., {Dowell}, C.~D., {Lim}, T., {et~al.} 2008, in SPIE Conference
  Series, Vol. 7010

\bibitem[{{Hailey-Dunsheath} {et~al.}(2010){Hailey-Dunsheath}, {Nikola},
  {Stacey}, {Oberst}, {Parshley}, {Benford}, {Staguhn}, \& {Tucker}}]{hd10}
{Hailey-Dunsheath}, S., {Nikola}, T., {Stacey}, G.~J., {et~al.} 2010, \apjl,
  714, L162

\bibitem[{{Hainline} {et~al.}(2009){Hainline}, {Blain}, {Smail}, {Frayer},
  {Chapman}, {Ivison}, \& {Alexander}}]{hainline09}
{Hainline}, L.~J., {Blain}, A.~W., {Smail}, I., {et~al.} 2009, \apj, 699, 1610

\bibitem[{{Holland} {et~al.}(2006){Holland}, {MacIntosh}, {Fairley}, {Kelly},
  {Montgomery}, {Gostick}, {Atad-Ettedgui}, {Ellis}, {Robson}, {Hollister},
  {Woodcraft}, {Ade}, {Walker}, {Irwin}, {Hilton}, {Duncan}, {Reintsema},
  {Walton}, {Parkes}, {Dunare}, {Fich}, {Kycia}, {Halpern}, {Scott}, {Gibb},
  {Molnar}, {Chapin}, {Bintley}, {Craig}, {Chylek}, {Jenness}, {Economou}, \&
  {Davis}}]{holland06}
{Holland}, W., {MacIntosh}, M., {Fairley}, A., {et~al.} 2006, in SPIE
  Conference Series, Vol. 6275

\bibitem[{{Hollenbach} \& {Tielens}(1999)}]{hollenbach99}
{Hollenbach}, D.~J. \& {Tielens}, A.~G.~G.~M. 1999, Reviews of Modern Physics,
  71, 173

\bibitem[{{Ivison} {et~al.}(2010{\natexlab{a}}){Ivison}, {Magnelli}, {Ibar},
  {Andreani}, {Elbaz}, {Elbaz}, {Elbaz}, {Andreani}, {Elbaz}, {Elbaz}, {Elbaz},
  {Andreani}, {Elbaz}, {Elbaz}, \& {Elbaz}}]{ivison10a}
{Ivison}, R.~J., {Magnelli}, B., {Ibar}, E., {et~al.} 2010{\natexlab{a}}, \aap,
  this volume

\bibitem[{{Ivison} {et~al.}(1998){Ivison}, {Smail}, {Le Borgne}, {Blain},
  {Kneib}, {Bezecourt}, {Kerr}, \& {Davies}}]{ivison98}
{Ivison}, R.~J., {Smail}, I., {Le Borgne}, J., {et~al.} 1998, \mnras, 298, 583

\bibitem[{{Ivison} {et~al.}(2010{\natexlab{b}}){Ivison}, {Smail},
  {Papadopoulos}, {Wold}, {Richard}, {Swinbank}, {Kneib}, \&
  {Owen}}]{ivison10b}
{Ivison}, R.~J., {Smail}, I., {Papadopoulos}, P.~P., {et~al.}
  2010{\natexlab{b}}, \mnras, 404, 198

\bibitem[{{Kaufman} {et~al.}(1999){Kaufman}, {Wolfire}, {Hollenbach}, \&
  {Luhman}}]{kaufman99}
{Kaufman}, M.~J., {Wolfire}, M.~G., {Hollenbach}, D.~J., \& {Luhman}, M.~L.
  1999, \apj, 527, 795

\bibitem[{{Kennicutt}(1998)}]{kennicutt98a}
{Kennicutt}, Jr., R.~C. 1998, \araa, 36, 189

\bibitem[{{Lord} {et~al.}(1996){Lord}, {Malhotra}, {Lim}, {Helou}, {Rubin},
  {Stacey}, {Hollenbach}, {Werner}, {Thronson}, {Beichman}, {Dinerstein},
  {Hunter}, {Lo}, \& {Lu}}]{lord96}
{Lord}, S.~D., {Malhotra}, S., {Lim}, T., {et~al.} 1996, \aap, 315, L117

\bibitem[{{Madden} {et~al.}(1993){Madden}, {Geis}, {Genzel}, {Herrmann},
  {Jackson}, {Poglitsch}, {Stacey}, \& {Townes}}]{madden93}
{Madden}, S.~C., {Geis}, N., {Genzel}, R., {et~al.} 1993, \apj, 407, 579

\bibitem[{{Men{\'e}ndez-Delmestre} {et~al.}(2009){Men{\'e}ndez-Delmestre},
  {Blain}, {Smail}, {Alexander}, {Chapman}, {Armus}, {Frayer}, {Ivison}, \&
  {Teplitz}}]{md09}
{Men{\'e}ndez-Delmestre}, K., {Blain}, A.~W., {Smail}, I., {et~al.} 2009, \apj,
  699, 667

\bibitem[{{Pilbratt}(et al.\ 2010)}]{pilbratt10}
{Pilbratt}, G. et al.\ 2010, \aap, this volume

\bibitem[{{Siringo} {et~al.}(2009){Siringo}, {Kreysa}, {Kov{\'a}cs},
  {Schuller}, {Wei{\ss}}, {Esch}, {Gem{\"u}nd}, {Jethava}, {Lundershausen},
  {Colin}, {G{\"u}sten}, {Menten}, {Beelen}, {Bertoldi}, {Beeman}, \&
  {Haller}}]{siringo09}
{Siringo}, G., {Kreysa}, E., {Kov{\'a}cs}, A., {et~al.} 2009, \aap, 497, 945

\bibitem[{{Smail} {et~al.}(1997){Smail}, {Ivison}, \& {Blain}}]{smail97}
{Smail}, I., {Ivison}, R.~J., \& {Blain}, A.~W. 1997, \apjl, 490, L5+

\bibitem[{{Stacey} {et~al.}(1991){Stacey}, {Geis}, {Genzel}, {Lugten},
  {Poglitsch}, {Sternberg}, \& {Townes}}]{stacey91}
{Stacey}, G.~J., {Geis}, N., {Genzel}, R., {et~al.} 1991, \apj, 373, 423

\bibitem[{{Swinbank} {et~al.}(2010){Swinbank}, {Smail}, {Longmore}, {Harris},
  {Baker}, {De Breuck}, {Richard}, {Edge}, {Ivison}, {Blundell}, {Coppin},
  {Cox}, {Gurwell}, {Hainline}, {Krips}, {Lundgren}, {Neri}, {Siana},
  {Siringo}, {Stark}, {Wilner}, \& {Younger}}]{swinbank10a}
{Swinbank}, A.~M., {Smail}, I., {Longmore}, S., {et~al.} 2010, \nat, 464, 733

\bibitem[{{Swinyard} {et~al.}(2010){Swinyard}, {Griffin}, {Ade}, {Ferkin},
  {Swinyard}, \& {Swinyard}}]{swinyard10}
{Swinyard}, B., {Griffin}, M., {Ade}, P., {et~al.} 2010, \aap, this volume

\bibitem[{{Tacconi} {et~al.}(2008){Tacconi}, {Genzel}, {Smail}, {Neri},
  {Chapman}, {Ivison}, {Blain}, {Cox}, {Omont}, {Bertoldi}, {Greve},
  {F{\"o}rster Schreiber}, {Genel}, {Lutz}, {Swinbank}, {Shapley}, {Erb},
  {Cimatti}, {Daddi}, \& {Baker}}]{tacconi08}
{Tacconi}, L.~J., {Genzel}, R., {Smail}, I., {et~al.} 2008, \apj, 680, 246

\bibitem[{{Vieira} {et~al.}(2010){Vieira}, {Crawford}, {Switzer}, {Ade},
  {Aird}, {Ashby}, {Benson}, {Bleem}, {Brodwin}, {Carlstrom}, {Chang}, {Cho},
  {Crites}, {de Haan}, {Dobbs}, {Everett}, {George}, {Gladders}, {Hall},
  {Halverson}, {High}, {Holder}, {Holzapfel}, {Hrubes}, {Joy}, {Keisler},
  {Knox}, {Lee}, {Leitch}, {Lueker}, {Marrone}, {McIntyre}, {McMahon}, {Mehl},
  {Meyer}, {Mohr}, {Montroy}, {Padin}, {Plagge}, {Pryke}, {Reichardt}, {Ruhl},
  {Schaffer}, {Shaw}, {Shirokoff}, {Spieler}, {Stalder}, {Staniszewski},
  {Stark}, {Vanderlinde}, {Walsh}, {Williamson}, {Yang}, {Zahn}, \&
  {Zenteno}}]{vieira10}
{Vieira}, J.~D., {Crawford}, T., {Switzer}, E., {et~al.} 2010, arXiv:0912.2338

\bibitem[{{Wei{\ss}} {et~al.}(2009){Wei{\ss}}, {Ivison}, {Downes}, {Walter},
  {Cirasuolo}, \& {Menten}}]{weiss09}
{Wei{\ss}}, A., {Ivison}, R.~J., {Downes}, D., {et~al.} 2009, \apjl, 705, L45

\bibitem[{{Wolfire} {et~al.}(1990){Wolfire}, {Tielens}, \&
  {Hollenbach}}]{wolfire90}
{Wolfire}, M.~G., {Tielens}, A., \& {Hollenbach}, D. 1990, \apj, 358, 116

\bibitem[{{Younger} {et~al.}(2008){Younger}, {Fazio}, {Wilner}, {Ashby},
  {Blundell}, {Gurwell}, {Huang}, {Iono}, {Peck}, {Petitpas}, {Scott},
  {Wilson}, \& {Yun}}]{younger08}
{Younger}, J.~D., {Fazio}, G.~G., {Wilner}, D.~J., {et~al.} 2008, \apj, 688, 59

\end{thebibliography}

\end{document}